\documentclass[aps,superscriptaddress]{revtex4-1}
\usepackage{epsfig}
\usepackage[colorlinks=true,citecolor=blue,urlcolor=blue,linkcolor=blue]{hyperref}
\usepackage{natbib}
\usepackage{times}
\usepackage{float}
\usepackage[normalem]{ulem} 

\begin{document}
\title{First observation with global network of optical atomic
clocks aimed for a dark matter detection}

\author{P. Wcis{\l}o}
\thanks{present affiliation: JILA, National Institute of Standards and Technology and the University of Colorado, Department of Physics, University of Colorado, Boulder, Colorado 80309-0440, USA
\affiliation{Institute of Physics, Faculty of Physics, Astronomy and Informatics, Nicolaus Copernicus University, Grudzi\c{a}dzka 5, PL-87-100 Toru\'n, Poland}
}
\author{P. Ablewski}
\affiliation{Institute of Physics, Faculty of Physics, Astronomy and Informatics, Nicolaus Copernicus University, Grudzi\c{a}dzka 5, PL-87-100 Toru\'n, Poland}
\author{K. Beloy}
\affiliation{National Institute of Standards and Technology, 325 Broadway, Boulder, CO 80305, USA}
\author{S. Bilicki}
\affiliation{Institute of Physics, Faculty of Physics, Astronomy and Informatics, Nicolaus Copernicus University, Grudzi\c{a}dzka 5, PL-87-100 Toru\'n, Poland}
\affiliation{LNE-SYRTE, Observatoire de Paris, Universit\'e PSL, CNRS, Sorbonne Universit\'e, 61 avenue de l'Observatoire 75014 Paris, France}
\author{M. Bober}
\affiliation{Institute of Physics, Faculty of Physics, Astronomy and Informatics, Nicolaus Copernicus University, Grudzi\c{a}dzka 5, PL-87-100 Toru\'n, Poland}
\author{R. Brown}
\affiliation{National Institute of Standards and Technology, 325 Broadway, Boulder, CO 80305, USA}
\author{R. Fasano}
\affiliation{National Institute of Standards and Technology, 325 Broadway, Boulder, CO 80305, USA}
\author{R. Ciury\l{}o}
\affiliation{Institute of Physics, Faculty of Physics, Astronomy and Informatics, Nicolaus Copernicus University, Grudzi\c{a}dzka 5, PL-87-100 Toru\'n, Poland}
\author{H. Hachisu}
\affiliation{National Institute of Information and Communications Technology, 4-2-1 Nukuikitamachi, Koganei, 184-8795 Tokyo, Japan}
\author{T. Ido}
\affiliation{National Institute of Information and Communications Technology, 4-2-1 Nukuikitamachi, Koganei, 184-8795 Tokyo, Japan}
\author{J. Lodewyck}
\affiliation{LNE-SYRTE, Observatoire de Paris, Universit\'e PSL, CNRS, Sorbonne Universit\'e, 61 avenue de l'Observatoire 75014 Paris, France}
\author{A. Ludlow}
\affiliation{National Institute of Standards and Technology, 325 Broadway, Boulder, CO 80305, USA}
\author{W. McGrew}
\affiliation{National Institute of Standards and Technology, 325 Broadway, Boulder, CO 80305, USA}
\author{P. Morzy\'nski}
\affiliation{Institute of Physics, Faculty of Physics, Astronomy and Informatics, Nicolaus Copernicus University, Grudzi\c{a}dzka 5, PL-87-100 Toru\'n, Poland}
\affiliation{National Institute of Information and Communications Technology, 4-2-1 Nukuikitamachi, Koganei, 184-8795 Tokyo, Japan}
\author{D. Nicolodi}
\affiliation{National Institute of Standards and Technology, 325 Broadway, Boulder, CO 80305, USA}
\author{M. Schioppo}
\affiliation{National Institute of Standards and Technology, 325 Broadway, Boulder, CO 80305, USA}
\affiliation{National Physical Laboratory (NPL), Teddington, TW11 0LW, United Kingdom}
\author{M. Sekido}
\affiliation{National Institute of Information and Communications Technology, 4-2-1 Nukuikitamachi, Koganei, 184-8795 Tokyo, Japan}
\author{R. Le~Targat}
\affiliation{LNE-SYRTE, Observatoire de Paris, Universit\'e PSL, CNRS, Sorbonne Universit\'e, 61 avenue de l'Observatoire 75014 Paris, France}
\author{P. Wolf}
\affiliation{LNE-SYRTE, Observatoire de Paris, Universit\'e PSL, CNRS, Sorbonne Universit\'e, 61 avenue de l'Observatoire 75014 Paris, France}
\author{X. Zhang}
\affiliation{National Institute of Standards and Technology, 325 Broadway, Boulder, CO 80305, USA}
\author{B. Zjawin}
\affiliation{Institute of Physics, Faculty of Physics, Astronomy and Informatics, Nicolaus Copernicus University, Grudzi\c{a}dzka 5, PL-87-100 Toru\'n, Poland}
\author{M. Zawada}
\email[]{zawada@fizyka.umk.pl}
\affiliation{Institute of Physics, Faculty of Physics, Astronomy and Informatics, Nicolaus Copernicus University, Grudzi\c{a}dzka 5, PL-87-100 Toru\'n, Poland}

\date{\today}

\begin{abstract}
We report on the first earth-scale quantum sensor network based on optical atomic clocks aimed at dark matter (DM) detection. Exploiting differences in the susceptibilities to the fine-structure constant of essential parts of an optical atomic clock, i.e. the cold atoms and the optical reference cavity, we can perform sensitive searches for dark matter signatures without the need of real-time comparisons of the clocks. We report a two orders of magnitude improvement in constraints on transient variations of the fine-structure constant, which considerably improves the detection limit for the standard model (SM) - DM coupling. We use Yb and Sr optical atomic clocks at four laboratories on three continents to search for both topological defect (TD) and massive scalar field candidates. No signal consistent with a dark-matter coupling is identified, leading to significantly improved constraints on the DM-SM couplings.
\end{abstract}

\maketitle

\section{Introduction}

Evidence for the existence of  dark matter comes entirely from astrophysical observations at large scales (galactic to cosmological).  The nature of  DM composition, however, will be  known only after positive detection of the DM candidates. The viable cold DM candidates, such as axions  \cite{Wilczek1978-va,Weinberg1978-qr,Olive2016-nh},  WIMPs \cite{Olive2016-nh},  and super-WIMPs \cite{Feng2003-eh}, require the existence of fields that can be coupled to standard model fields. Therefore, existing experiments focus on searches for such couplings, though no experimental data has provided a positive detection. In the LUX experiment, for example, which studied potential coupling between WIMPs and nucleons, constraints on the scattering cross section per nucleon were reported below $10^{-45}$~cm${}^2$ \cite{noauthor_2014-um}. New stringent upper limits were recently reported in the (8-125) keV/c${}^2$  mass range that excluded couplings to electrons with coupling constants of $g_{ae}>3\times 10^{-13}$ for pseudo-scalar and $\alpha'>2\times 10^{-28}\times\alpha$ ($\alpha$ is the fine structure constant, and $\alpha'$ is its vector  boson equivalent) for vector super-WIMPs, respectively, by the XENON100 experiment \cite{noauthor_2017-zy}.  

Lack of any detected DM in the form of particles yielded alternative theories, such as oscillating massive scalar fields  \cite{Arvanitaki2015-bk} or topological defects in the scalar fields \cite{Derevianko2014-tz,Wcislo2016-cu}. The coupling of the scalar field $\phi$ to the SM fermion and electromagnetic fields may be represented by linear or quadratic interactions \cite{Damour2010-iu,Stadnik2016-ej,Stadnik2014-rx} and could be detected by observing possible variations of the effective fine-structure constant \cite{Derevianko2014-tz,Wcislo2016-cu,Hees2016-cf,Van_Tilburg2015-uf,Kalaydzhyan2017-wi,Roberts2017-xk}. 

Exceptional accuracy and stability of optical atomic clocks and ultra-narrow lasers open the door for new tests of fundamental physics that can reach beyond the Standard Model. Optical atomic clocks were used, for example, to constrain the coupling of fundamental constants such as the fine-structure constant $\alpha$, electron-proton mass ratio $\mu$, and light quark mass to gravity~\cite{Blatt2008}. Optical atomic clocks connected by phase compensated optical fibre links improved a constraint on the Robertson-Mansouri-Sexl parameter quantifying a violation of time dilation~\cite{Delva2017}. Stability of optical clocks and ultra-narrow lasers in our network allows for determination of a new constraint on the transient perturbation in the fine-structure constant $\left\vert\delta\alpha/\alpha\right\vert <1.6 \times 10^{-16}$ which is an improvement by two orders of magnitude.

Recently, we have shown that a single optical atomic clock can be used as a detector for  dark matter in the form of stable topological defects \cite{Wcislo2016-cu}. This is possible by taking advantage of the differences in the susceptibilities to the fine structure constant between essential parts of any optical atomic clock, i.e. the atoms and the cavity used to pre-stabilize the clock laser.  Here, we realize  the first earth-scale quantum sensor network based on optical atomic clocks. In our approach, measurements by the distant clocks do not need to be linked in real time. In analogy to the standard radio-astronomical technique, very-long-baseline interferometry, clock measurements  can be locally recorded (with time stamps that are accurate to the level of 1~ms) and cross-correlated later. A concept of archiving data from various types of precise measurements was also proposed in \cite{Budker2015}. In this letter, we report the results of measurements of  Yb and Sr optical clocks at four laboratories and on three continents to search for topological defect and massive scalar field candidates.

\section{Results}

\begin{figure}[h]
	\centering
	\includegraphics[width=1\columnwidth]{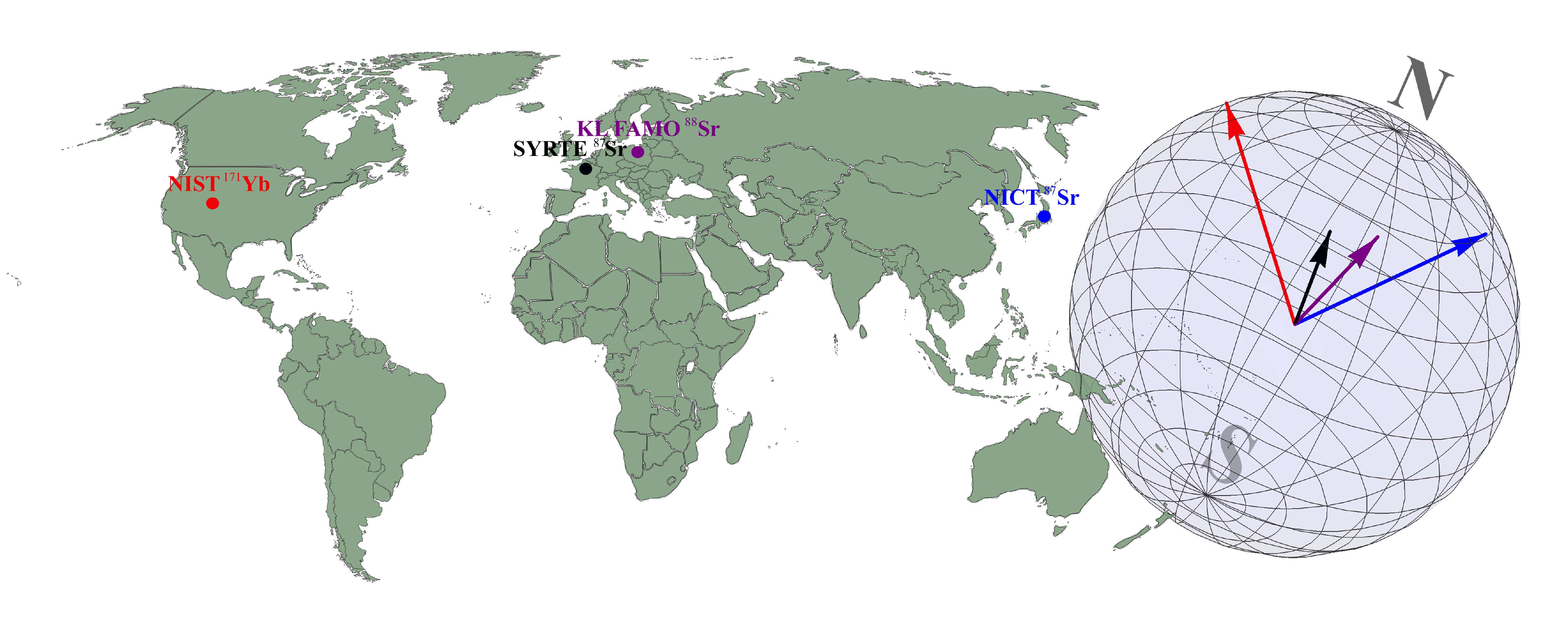} 
	\caption{\textbf{ Global sensor network.} The participating optical lattice atomic  clocks reside  at NIST, Boulder, CO, USA \cite{Schioppo2016-va,Hinkley2013-su}, LNE-SYRTE, Paris, France \cite{Le_Targat2013-eu,Lodewyck2016-pk},  KL FAMO, Torun, Poland \cite{Morzynski2015-yn,Bober2015-up}, and NICT, Tokyo, Japan \cite{Hachisu2015-kq,Hachisu2018}. 
    %The vector position of the laboratories used in the analysis are  acquired form the International Terrestrial Reference Frame (ITRF) database that provide ITRF coordinates in requested frame and epoch.
    }
	\label{fig:world}
\end{figure}

Our network is composed of optical lattice atomic clocks located at NIST, Boulder, CO, USA \cite{Schioppo2016-va,Hinkley2013-su}, at LNE-SYRTE, Paris, France \cite{Le_Targat2013-eu,Lodewyck2016-pk}, at KL FAMO, Torun, Poland \cite{Morzynski2015-yn,Bober2015-up}, and at NICT, Tokyo, Japan \cite{Hachisu2015-kq,Hachisu2018} (see Fig.~\ref{fig:world}). Each clock utilizes an optical-cavity-stabilized laser that is frequency tuned to resonantly interrogate the atomic clock transition of cold atoms trapped in an optical lattice.   The clock records the local difference $r_{i}(t)$  between the atomic transition frequency, $\nu_{atom}$, and the cavity-stabilized-laser, $\nu_{cavity}$, where $i=$1\ldots4 corresponds to the four clocks in the network. Any variation in the electromagnetic fine-structure constant, $\alpha$, will be imprinted in this difference because of different susceptibilities of these two frequencies to variations in $\alpha$ ($\nu_{atom}/\nu_{cavity}\propto \alpha$)  \cite{Stadnik2015-cq,Stadnik2016-cm,Wcislo2016-cu}.
%{\color{red}\sout{($\nu \propto \alpha^2$ for non-relativistic atoms and $\nu \propto \alpha$ for the cavity) in general the powers of these dependencies depend on the choice of the frequency unit, but the relative difference of these frequencies is not. Here we choose as a fixed frequency unit $\omega_{Compton}=m_e c^2/\hbar$.}}
In particular, we may expect variations in $\alpha$ that can be expressed as  $\delta\alpha/\alpha = \left(\phi/\Lambda_{\gamma,n}\right)^n$ where $\phi$ is the DM field and $\Lambda_{\gamma,n}$ is the energy scale (which inversely parametrizes the strength of the DM-SM coupling) of the n-th portal \cite{Essig_Rouven_and_Jaros_John_A_and_Wester_William_et_al2013-mm}.

The general potential of self-interaction of the scalar field  can yield an evolution of the  scalar field  in the form of a coherently oscillating classical field \cite{Hees2016-cf}, as well as  spontaneous symmetry breaking and, as a consequence, to the formation of topological solitons, also called topological defects  \cite{Vilenkin1985-lb}. Both oscillating field and topological defects were recently proposed  as possible forms of the DM \cite{Arvanitaki2015-bk,Derevianko2014-tz}, and may both induce a variation of the fine-structure constant in the form of oscillations or transient changes, respectively \cite{Derevianko2016-pq}. The linear coupling n = 1 is usually considered in the oscillating massive scalar field DM studies \cite{Damour2010-mr,Arvanitaki2015-bk,Hees2016-cf}, while quadratic coupling n = 2 is usually considered for the topological DM  \cite{Stadnik2015-ho,Derevianko2014-tz,Wcislo2016-cu}.  One can also translate limits from linear to quadratic couplings \cite{Stadnik2016-ej}. 

\begin{figure}[h]
	\centering
	\includegraphics[width=\columnwidth]{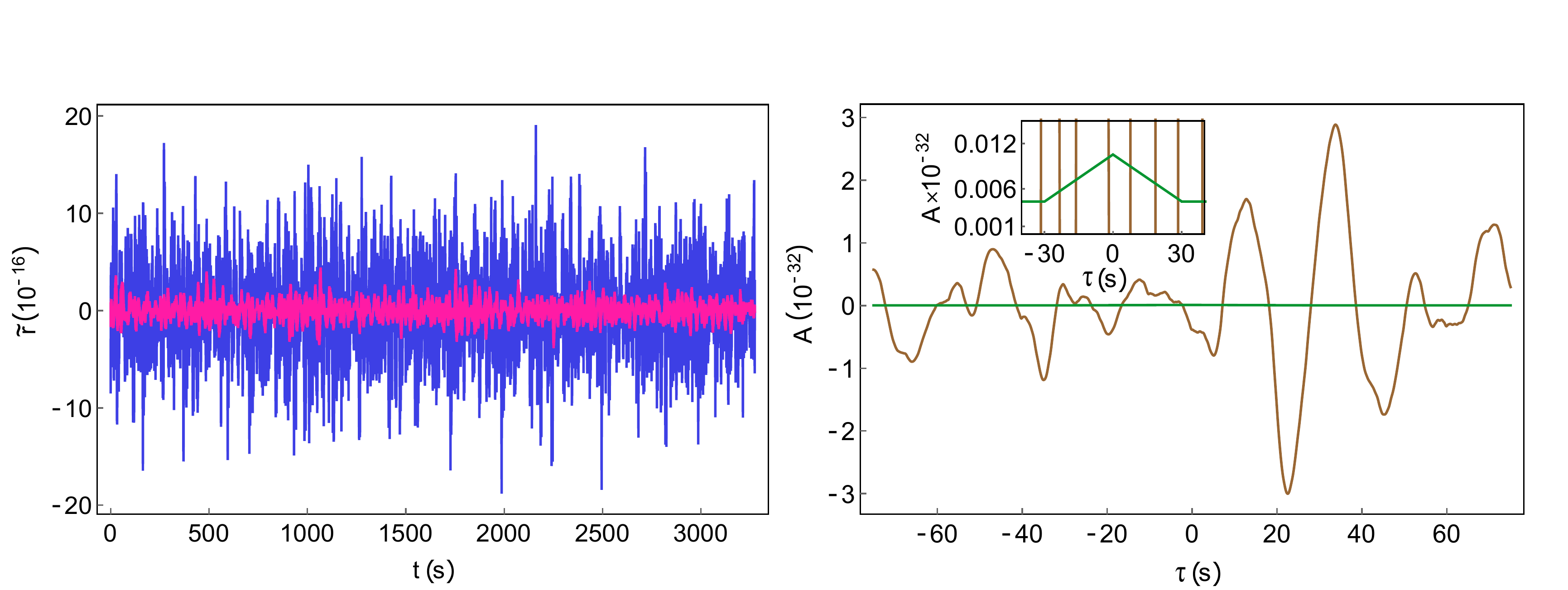}
	\caption{{\bf Cross-correlation of two $\tilde{r}_i(t)$.} Left panel: Overlapping normalized signals, $\tilde{r}_i(t)$, from the NIST and LNE-SYRTE laboratories (pink and blue lines, respectively).
Right: cross-correlation, $A(\tau)$, of the two $\tilde{r}_i(t)$ (brown line) and the fit of the expected cross-correlation function assuming a square perturbation with duration of 30~s (green line). $\tau$ is the cross-correlation displacement. The inset depicts the magnification of the region where the fitted triangular function can be seen.}
\label{fig:raw} 
\end{figure}

In the case of topological defects, the transient effect can be detected by cross-correlating the pairs of $r_i(t)$ recorded in our geographically dispersed clocks. To allow for the comparison between  clocks based on different atomic species and different transitions, we define the normalized signals as $\tilde{r}_i(t)=r_i(t)/(K_\alpha^i \nu_0^i)$, where $K_\alpha^i$ is the sensitivity coefficient (equal to one in the non-relativistic case \cite{Stadnik2015-cq,Stadnik2016-cm}) and $\nu_0^i$ is the frequency of the clock transition. A smooth high-pass FIR (finite impulse response) filter with frequency cut-off between 0.005 and 0.027~Hz (depending on the size of the sought defect) sampling both past and future points weighted with a Gaussian profile is applied to all $\tilde{r}_i(t)$ to eliminate the low-frequency drifts that originate from the long-term cavities' instabilities. We model the transient effect by a square perturbation of the fine-structure constant with size $d$, and with an amplitude assumed to be identical for the distant clocks of our network. 
We take into account rotation and revolution of Earth around the Sun as well as revolution of our Solar System around the center of our galaxy by assuming  the  topological defects' speed relative to  Earth $v = 300$ km~s$^{-1}$ and considering all possible direction of the topological defects' relative velocity.
We validated this model by showing that, for the most constrained limit, the results calculated with a square profile  differ only by a few percent from the results calculated with a simple model of a stable plain domain wall in the $xy$-plane with  transverse profile of a scalar field  $\phi(z)\sim tanh(z/L)$, where $L \approx d/2$ is the thickness of the wall  orthogonal to the direction $z$ of defect propagation \cite{Vilenkin1985-lb,Ya_B_Zeldovich_I_Yu_Kobzarev_LB_Okun1975-ju}. Our analysis was performed for the defect size, $d$, between $3\times 10^2$ and $3\times 10^4$ km.
% For perturbations of the fine-structure constant longer than one clock cycle $T_c$, (i.e. $d / v > T_c$,  where $v$ is the  topological defects' speed relative to the Earth, assumed to be $v = 300$ km~s$^{-1}$), 
We cross-correlate the pairs of $\tilde{r}_i(t)$ and  fit the resulting data with the expected DM signature (see 'Methods').
The left panel in Fig.~\ref{fig:raw} shows two $\tilde{r}_i(t)$ from NIST and LNE-SYRTE laboratories. The right panel shows their cross correlation and the fit of the expected shape (triangle) for the 30~s-long event. The inset presents a fragment of the fitted function with triangular shape.

\begin{figure}[H]
	\centering
	\includegraphics[width=0.8\columnwidth]{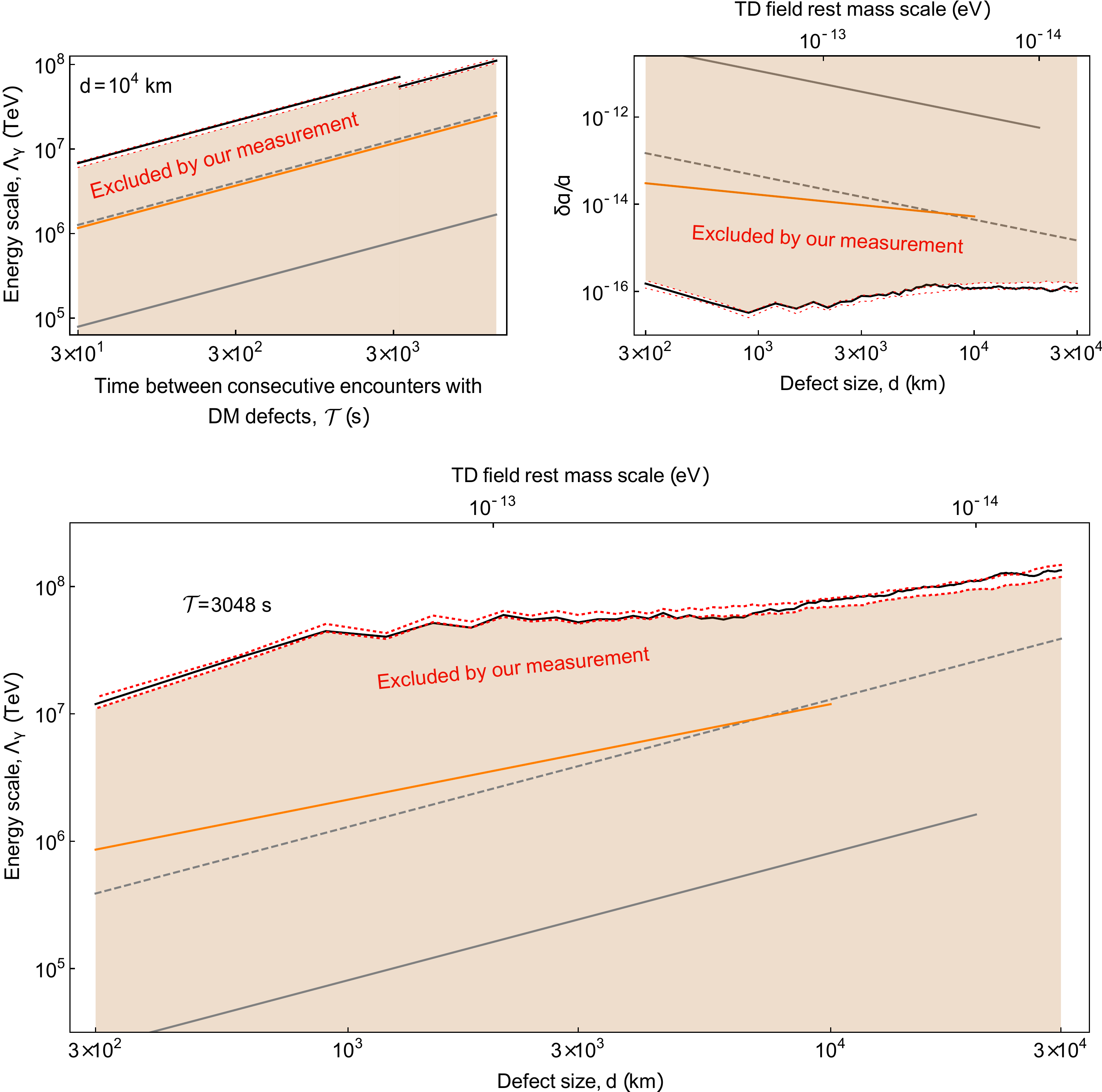}
	\caption{\textbf{ Constraints on the energy scale  $\Lambda_{\gamma,2}$ and the fine-structure variation $\delta\alpha/\alpha$.}
The top right: the constraints on the fine-structure variation $\delta\alpha/\alpha$ (solid black line). The top left and bottom panels: the constraints on the energy scale  $\Lambda_{\gamma,2}$ under the assumption of the mean cold DM energy density $\rho_{DM} = 0.4$~GeV~cm$^{-3}$ \cite{Olive2016-nh} (solid black line).
	  The orange lines represent the previous best constraints measured with two co-located ${}^{88}$Sr clocks \cite{Wcislo2016-cu}, and the grey solid and dashed lines represents the actual and possible  constraints, respectively, estimated from the GPS constellation's on-board clocks \cite{Kalaydzhyan2017-wi,Roberts2017-xk}.
     % , and the light and dark green lines are the future  projected sensitivities of the microwave clocks aboard the International Space Station, in ACES and SOC missions, respectively \cite{Kalaydzhyan2017-wi}. 
      The topological defect size, $d$, can be expressed as the TD field mass scale $m_{TD}= \hbar / (d c)$. The constraints in the top left panel have a discontinuity at ${\cal T}\approx 3000$~s because the $r_i(t)$ from the most stable clock in our network (NIST, Boulder, CO, USA) was collected over a shorter duration. The red dotted curves depict the   5\% and 95\% confidence levels.}
	\label{fig:td}
\end{figure}

 The constraint on the transient variation of the fine-structure constant can be expressed as:

\begin{equation}
\left\vert\frac{\delta\alpha}{\alpha}\right\vert < \sqrt{{A_0}/{\eta_T}}, \label{eq:deltaalpha}
\end{equation}

\noindent
where $A_0$ is the fitted amplitude of the cross-correlation peak and $\eta_T$ is the ratio of the overall defect duration to the duration of the cross-correlated data. 
The top right panel of Fig.~\ref{fig:td} shows the bound on the transient variation in the fine-structure, $\delta\alpha/\alpha$, derived from our measurement. The black line comes from the direct fit of the expected DM signature to the cross correlation. To give a proper statistical interpretation of our result and to have an ability to distinguish the noise from any exotic physical signal we calculate also the 5\% and 95\% confidence levels (CLs). To do it we artificially shift one of the readouts by a large period ensuring that no physical correlation between the two channels should be expected. We repeat this procedure 100 times for different shifts. For the particular case of  a defect size comparable to the size of the Earth ($d = 10 000$ km), we determine a new constraint  on the transient perturbation in the fine-structure constant:

\begin{equation}
\left\vert\frac{\delta\alpha}{\alpha}\right\vert < 1.6 \times 10^{-16},
\end{equation}

\noindent
which is almost two orders of magnitude better than the best limit published so far  \cite{Wcislo2016-cu}, represented by the orange lines in Fig.~\ref{fig:td}. For other values of $d$ the improvement is even larger and approaches 3 orders of magnitude for $d=900$~km. For comparison, we also plotted the actual and possible limits derived from the GPS constellation's on-board clocks (solid and dashed  gray lines, respectively) \cite{Roberts2017-xk}.
%,  and the future projected sensitivities of the microwave clocks aboard the International Space Station, in ACES and SOC missions (the light and dark green lines respectively)\cite{Kalaydzhyan2017-wi}. 

Equation (\ref{eq:deltaalpha}) can be translated into a lower bound for the energy scale $\Lambda_{\gamma,2}$  of the quadratic DM-SM coupling:

\begin{equation}
\Lambda_{\gamma,2} > d^{1 / 2}\sqrt{\sqrt{\frac{\eta_T}{A_0}}\rho_{DM} \hbar c {\cal T} v },\label{eq:lambda}
\end{equation}

\noindent
where $\rho_{DM}$ is the mean cold DM energy density in our local galactic neighbourhood, 
%$v$ is the  topological defects' speed relative to  Earth, assumed to be $v = 300$ km~s$^{-1}$,
and ${\cal T}$  is  the time between consecutive encounters with topological defects.  We assume a plane wave model of TD.
 Under the assumption of $\rho_{DM} = 0.4$~GeV~cm$^{-3}$ \cite{Olive2016-nh}, we derived the constraints on the energy scale  $\Lambda_{\gamma,2}$, plotted in the top left and bottom panels of  Fig.~\ref{fig:td} as a function of   ${\cal T}$ and $d$.
 The constraints in the top left panel have a visible discontinuity at ${\cal T}\approx 3000$~s. Our sensitivity for longer times between consecutive encounters ${\cal T}$ is lower  because the $r_i(t)$ from the most stable clock in our network (NIST, Boulder, CO, USA) was collected over a shorter duration. 
%For comparison, we also plotted in the low right panel in Fig.~\ref{fig:td} the future projected sensitivities of the microwave clocks aboard the International Space Station, in ACES and SOC missions (the light and dark green lines respectively)\cite{Kalaydzhyan2017-wi}. 

In the case of oscillating massive scalar fields,  periodic perturbations will occur with the same phase in all $r_i(t)$ signals.  The strength of the linear coupling is characterised in the literature \cite{Damour2010-iu,Damour2010-mr,Van_Tilburg2015-uf,Hees2016-cf} by the appropriate dimensionless coefficient:

\begin{equation}
d_e = \frac{M_P c^2}{\sqrt{4\pi}\Lambda_{\gamma,1}},
\end{equation}

\noindent
where $M_P c^2=1.2 \times 10^{19}$ GeV is the Planck mass energy equivalent \cite{Kalaydzhyan2017-wi}. 

\begin{figure}[h]
	\centering
	\includegraphics[width=0.8\columnwidth]{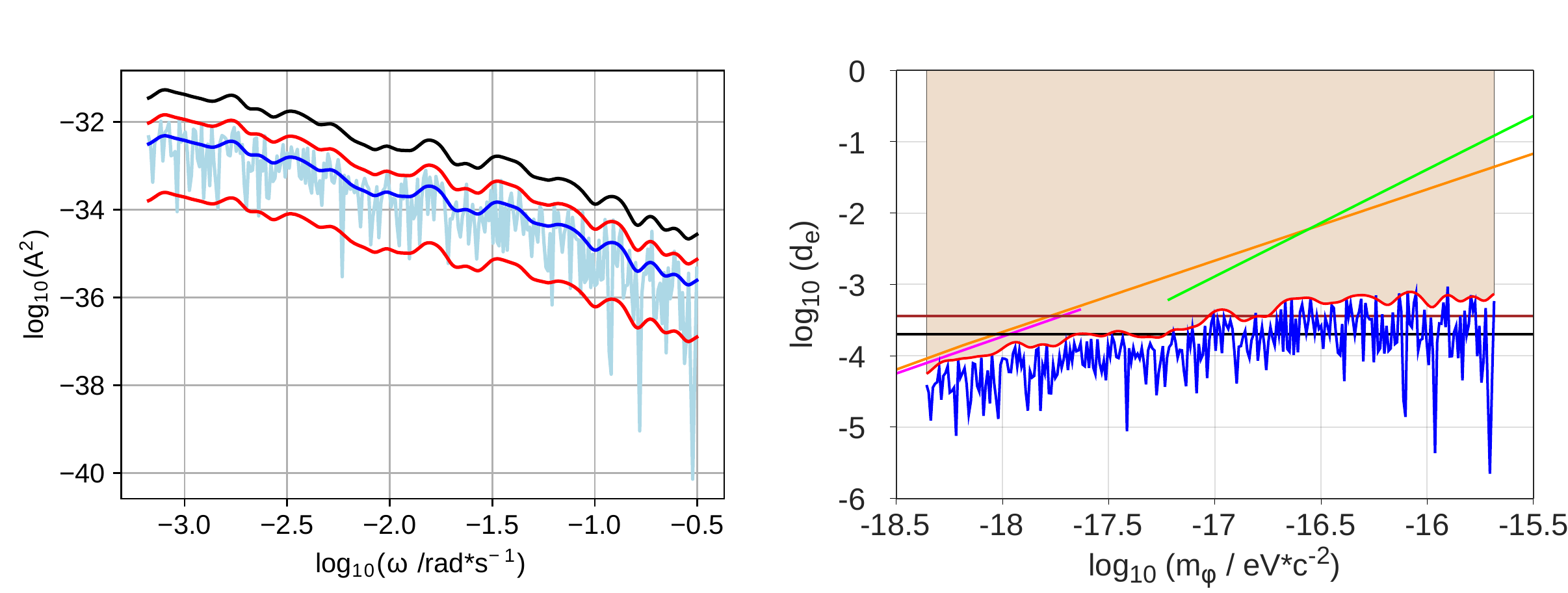}
	\caption{\textbf{ Upper limits on the coupling constant $d_e$.}   Left: periodogram of the square of the fitted amplitude $A(\omega)$ (blue). The $A^2(\omega)$ distribution in the data is consistent with pink (flicker) noise~\cite{Groth1975,Scargle1982-xm}. Lower and upper red curves depict the  5\% and 95\% confidence levels, respectively. The dark blue curve corresponds to the mean value based on the assumed noise model. The black curve is the detection threshold. 
%	Right: the dark and light blue curves  depict the limit on $d_e$ derived from our measurement.  
		Right: the blue curve  depicts the limit on $d_e$ derived from our measurement.  
%	For  masses $m_\phi >10^{-15.4}$~eV~c${}^2$ (light blue curve) the coherence time of the field is shorter than the duration of our observations, and we split $\tilde{r}_i(t)$ into periods equal to half of the given coherence time and assumed independent phases in each of the periods.  
	The red curve depicts the 95\% confidence level. The magenta and orange lines are the 95\% confidence limits reported in  Refs. \cite{Hees2016-cf} and \cite{Van_Tilburg2015-uf}, respectively.  The green line is the result derived from the stability analysis of a single optical atomic clock \cite{Kalaydzhyan2017-wi}. The brown line represents the equivalence principle (EP) tests such as the E\"ot-Wash experiment  \cite{Schlamminger2008-pf} and Lunar Laser Ranging \cite{Williams2004-ou}. The black line represents the first results of the MICROSCOPE experiment \cite{Berge2017-ys}.}
	\label{fig:osc}
\end{figure}

We applied the procedure used in  \cite{Van_Tilburg2015-uf,Hees2016-cf} and described in \cite{Scargle1982-xm} to search for harmonic-oscillation signatures. For each oscillation frequency  $\omega$ (i.e. for each mass $m_\phi=\hbar\omega /c^2$) we fit  the normalised signals $\tilde{r}_i(t)$ from all our detectors with an oscillating function $R(t) = B+A(\omega) cos(\omega t + \delta)$ function, where $B$ is a constant offset (the contributions were weighted with inverse corrected sample standard deviation for each $\tilde{r}_i(t)$).  The fitted $A^2(\omega)$ and the corresponding 5\% and 95\% confidence levels as well as the threshold levels are shown in the left panel of Fig.~\ref{fig:osc}. 
The periodogram allows identification of the noise type dominant in our data \cite{Groth1975,Scargle1982-xm}. The dependence of the $A^2(\omega)$ on frequency is consistent  with pink (flicker) noise. The noise analysis used to determine the confidence and threshold levels is described in 'Methods'.

The corresponding limit on the strength of the DM-SM coupling can be calculated as (see details in Ref. \cite{Hees2016-cf})

\begin{equation}
 d_e <\sqrt{\frac{A^2(\omega)\omega^2 c^2 }{\rho_{DM}8\pi G}}, 
\end{equation}

\noindent
where G is Newton's constant.
The blue curve in the right panel in Fig.~\ref{fig:osc} depict the limit on $d_e$ derived from our measurements as a function of the scalar field mass $m_\phi$.  
%The dark and light blue curves in the right panel in Fig.~\ref{fig:osc} depict the limit on de derived from our measurements as a function of the scalar field mass $m_\phi$. For masses higher than $10^{-15.4}$~eV~c${}^2$ the coherence time of the field is shorter than the duration of our observations. In this range we split $\tilde{r}_i(t)$  into periods equal to half of the given coherence time, and we fit the $R(t)$ function with independent phases but with common amplitude $A(\omega)$ in each of the periods. 
The coherence time of the field is longer than the duration of our observations for masses lower than $10^{-15.4}$~eV~c${}^{-2}$.
In the considered range of $m_\phi$, our results reached the same levels as the limits reported in Ref. \cite{Hees2016-cf} (magenta line), and improve those of \cite{Van_Tilburg2015-uf} and \cite{Kalaydzhyan2017-wi} (red and green lines, respectively). The reported results for the oscillating massive scalar fields are complementary to the limits derived from the weak equivalence principle tests  \cite{Schlamminger2008-pf,Williams2004-ou,Berge2017-ys}.

\section{Conclusions}

For the first time, we have demonstrated  operation of an Earth-scale network of optical atomic clocks. Different susceptibilities to the fine-structure constant of the atoms and the cavity allowed the intercontinental comparison of different clocks without real-time frequency transfer. Our measurements aim at searching for oscillations and transient variations in the fine-structure constant, which may originate from the presence of  dark matter in the form of topological defects or massive scalar fields. We improve the previous constraints on transient variations in $\alpha$  by two orders of magnitude. The approach demonstrated here has a large potential for further improvements. Longer, coordinated measurement intervals, and the incorporation of other optical atomic clocks into the network will allow for further improvements in the detection limits as well as for applying the determined constraints to larger volumes in the parameter space.

\section{Methods}

\subsection{Data sets $r_i(t)$}

We use $r_i(t)$ from four optical lattice atomic clocks located in NIST, Boulder, CO, USA \cite{Schioppo2016-va,Hinkley2013-su}, in LNE-SYRTE, Paris, France \cite{Le_Targat2013-eu,Lodewyck2016-pk}, in KL FAMO, Torun, Poland \cite{Morzynski2015-yn,Bober2015-up} and in NICT, Tokyo, Japan \cite{Hachisu2015-kq,Hachisu2018}. During the measurement sessions, the reported fractional clock instabilities at 1~s were equal to $3\times10^{-16}$, $1\times10^{-15}$, $2\times10^{-14}$, $7\times10^{-15}$,  and the reported fractional optical cavity instabilities  at 1~s  were equal to 
$2\times10^{-16}$, $8\times10^{-16}$, $2\times10^{-14}$, $2\times10^{-15}$,  
 in NIST, LNE-SYRTE, KL FAMO, and NICT laboratories, respectively.
The measurements were performed from 9 January 2015 to 17 December 2015. The measurements were collected for 11, 24, 42 and 54 days in NIST, LNE-SYRTE, KL FAMO and NICT laboratories, respectively. Together the laboratories collected data for 114 days. In our topological defect analysis we use the longest consecutive overlapping records for each pair of clocks. Their length is equal to 3336 s and 13533 s for NIST and LNE-SYRTE, and for LNE-SYRTE and KL FAMO, respectively. It should be noted that the constraints reported in Fig.~\ref{fig:td} are meaningful only if they are referred to the lengths of the analyzed signal. In oscillating massive scalar fields analysis we use all our data spanning 114 days from all the laboratories. One clock cycle lasts 0.5 s in NIST, 1 s in LNE-SYRTE, 1.5~s in NICT, and 1.3~s in KL FAMO. 
The clock cycle duration restricts the size of the topological defects that we can detect with the network of unsynchronized clocks.
In the case of oscillating scalar fields the results are limited by the feedback servos' time constants, which varies from 2 to 20~s depending on the clock.

In our oscillating fields analysis we weight $r_i(t)$  from each clock with their standard deviations. 
All $r_i(t)$ are heavily affected by low-frequency cavity drifts. Before the analysis a high-pass filter is applied to eliminate the low-frequency cavities' instabilities. In the case of topological defects' analysis a FIR (finite impulse response) high-pass Gaussian filter with frequency cut-off at 0.005 to 0.027 Hz is applied to all $r_i(t)$. Such choice of transmission characteristics increases the sensitivity to the considered range of events.
In the oscillating massive scalar fields' analysis, the main contribution to the signal instabilities are cavity drifts in much longer time scale. We filter them by subtracting a second-order polynomial fit from continuous sections of $r_i(t)$. To verify the filter response, we add artificial harmonic oscillations with specified parameters to non-filtered data and we test how the filter performs on different frequencies.

\subsection{Topological defects: data analysis}

For nonzero DM-SM coupling, a topological defect will manifest as a transient perturbation of the fine-structure constant. We denote its magnitude and duration as $\delta\alpha$ and $T$, respectively. The corresponding perturbation of the relative cavity-atom frequency encoded in $r_i(t)$ depends on the atom sensitivity $K_\alpha$ and for two different clocks A and B can be written as

\begin{eqnarray}
r_A(t) & =K_\alpha^A \nu_0^A\frac{\delta\alpha}{\alpha},\nonumber\\
r_B(t) & =K_\alpha^B \nu_0^B\frac{\delta\alpha}{\alpha}.\nonumber\\
\end{eqnarray}

We define normalized signals as $\tilde{r}_i(t)=r_i(t)/(K_\alpha^i \nu_0^i)$. In the simple case involving two clocks, the goal of our data analysis is to determine the constraint on $\delta\alpha$ from $\tilde{r}_A(t)$ and $\tilde{r}_B(t)$. This can be done by cross correlating  $\tilde{r}_A(t)$ and $\tilde{r}_B(t)$

\begin{equation}
(\tilde{r}_A*\tilde{r}_B)(\tau)=\frac{1}{t_B-t_A}\int_{t_A}^{t_B}\tilde{r}_A(t)\tilde{r}_B(t+\tau)dt,
\end{equation}

\noindent
where  $t_B-t_A$ is the duration of the cross correlated signals. 
For square events the shape of their cross correlation is triangular with full width at half maximum equal to $T$ and amplitude 

\begin{equation}
A_0=\eta_T\left(\frac{\delta\alpha}{\alpha}\right)^2.
\end{equation}

\noindent
This gives the constraint on the transient variations in the fine-structure constant (see Eq. (\ref{eq:deltaalpha}) in the main text). For the quadratic coupling, $\delta\alpha/\alpha=\phi^2/\Lambda_{\gamma,2}^2$, Eq. (\ref{eq:lambda}) directly follows from Eq. (\ref{eq:deltaalpha}); note that the DM field inside a defect is $\phi^2={\cal T} v \rho_{DM}d\hbar c$ \cite{Derevianko2014-tz}. The DM field is assumed to be zero outside the TDs.

The simplest model of  DM topological defects requires at least two free parameters whose values have to be arbitrarily chosen; in our case these parameters are the defect size, $d$, and the time between consecutive encounters with topological defects, $\cal{T}$. We repeat our analysis for every combination of these two parameters within the ranges we access with our network. For given values of $d$ and $\cal{T}$, we cross correlate the two $\tilde{r}_i(t)$ and fit the cross correlation with the expected triangular shape. We repeat this fitting procedure for any accessible time and any time delay between the two labs. The maximal value of the fitted amplitude, $A_0$, yields the constraints on the transient variations in $\alpha$ (Eq. (\ref{eq:deltaalpha})), and strength of the DM-SM coupling (Eq. (\ref{eq:lambda})), see the black lines in Fig. \ref{fig:td}. This approach is analogous to the Very-Long-Baseline Interferometry (VLBI) techniques used in radio-astronomy. 

To give a proper statistical interpretation of our results we also calculate the 5\% and 95\% confidence levels (CLs), see the dashed red lines in Fig.~\ref{fig:td}. To calculate the CLs, we repeat (100 times) the above algorithm of maximal $A_0$ determination,  artificially shifting one of the $\tilde{r}_i(t)$ in time ensuring that no expected common physical signal is present (for every repetition we shift $\tilde{r}_i(t)$ by a different time). Having such ensemble of 100 cross-correlation amplitudes (for a given $d$) we select the five smallest and largest elements, which yields the 5\% and 95\% CLs. If we treat the 95\% CL as an actual constraint determined in this work then  the black line in the top right panel in Fig.~\ref{fig:td} should statistically (i.e. without any common component) exceed the the red dashed 95\% CL one time per every 20 values in $d$ (the step in $d$ is 300~km). This is consistent with  Fig.~\ref{fig:td}, which indicates that at present level of accuracy we do not observe any signature of hypothetical DM.

In the dataset considered here, the time of overlap of at least three clocks was considerably smaller than the time of overlap of two clocks. Therefore, without losing the strength of the constraint, we could restrict our analysis to simultaneous comparison of two clocks at given time. For more than two simultaneous $\tilde{r}_i(t)$, the approach presented here can be generalized in a straightforward way. For any value of the model free parameters and for any value of the DM velocity, the expected DM signature can be simultaneously fit to all the cross-correlations between the participating clocks. In our analysis we choose $\eta_T=1$, i.e. the length of the cross correlated $\tilde{r}_i(t)$  segments are equal to the event duration, which is optimal for this analysis. It should be mentioned that the sensitivity of the clocks to DM objects could drop by a factor of two for typical servo-loop settings when the DM object duration is comparable to one servo-loop cycle, $T_c$. 
%For even shorter duration the interpretation of our results is not straightforward. If we assume that during the observation time, a sufficiently large number of DM events occurs (randomly distributed in time), then our approach is still applicable. The probability that a short DM event will overlap with the interrogation period is smaller than one. However, the contribution of any such event to the height of the cross-correlation peak will be overestimated. These two effects cancel out and the sensitivity is the same as that for longer events, but the $\eta$ parameter is smaller than 1. 

\subsection{Oscillating massive scalar fields: data analysis}

For each frequency $\omega$, i.e. for each mass $m_\phi$, we perform the linear least-squares analysis of the normalised signal $\tilde{r}(t)$ composed of the $\tilde{r}_i(t)$ contributions from all the four laboratories.  We fit to the function $R(t) = B+A(\omega) cos(\omega t + \delta)$, where $B$ is a constant offset (the contributions from different labs are weighted with inverse variance). In the left panel in Fig.~\ref{fig:osc} the square of the fitted amplitude is shown as a function of $\omega$. To interpret the obtained result one has to verify if the fitted amplitude considerably exceeds the noise level. First let us consider the simplest case when the noise of $\tilde{r}(t)$ is white. The square amplitude of the harmonic component, $A^2(\omega)$, can be treated as a new random variable. For the case of white noise  ($N$ samples of normalized signal $\tilde{r}(t)$ with standard deviation $\sigma$) its expected value does not depend on $\omega$, $\langle A^2(\omega)\rangle=4\sigma^2/N$, and the cumulative probability distribution function can be expressed by a simple analytical formula \cite{Scargle1982-xm}

\begin{equation}
CDF(A^2)=1-e^{-A^2/\langle A^2\rangle}.
\end{equation}

\noindent
Then the $A^2$ corresponding to $X$ confidence level (CL) can be expressed as 

\begin{equation}
A^2_X=-\langle A^2\rangle ln(1-X). \label{eq:CLX}
\end{equation}

The statistical interpretation of Eq. (\ref{eq:CLX}) is that the probability that $A^2<A^2_X$ is $X$. For instance, for $X=95$\% the corresponding $A^2$ level is $A^2_X\approx2.996\langle A^2\rangle$.  Equation (\ref{eq:CLX}) concerns a single frequency (one out of the $N/2$ frequency channels available in the spectrum). In other words, one per 20 frequency channels on average should exceeds the $A_{95\%}^2$ level. Beyond the 95\% CL criterion we also define a detection threshold for $N_f$ frequency channels, $A^2_{DT}$, in a  similar way as  done in Refs. \cite{Scargle1982-xm,Hees2016-cf}: the random variable, $A^2$, exceeds the $A^2_{DT}$ level for any frequency channel, on average, only 0.05 times per measurement. In other words, if the measurement were repeated 100 times then the noise would be interpreted as a positive detection, on average, only five times. To make this definition unique we set additional condition that the probability of exceeding the detection threshold is the same for all the frequencies. For the case of white noise the detection threshold defined here can be expressed as 

\begin{equation}
A^2_{DT}=-\langle A^2\rangle ln\left(\frac{0.05}{N_f}\right).\label{eq:DT}
\end{equation}

\noindent
As a $N_f$ we take the total considered frequency range (see Fig. 3) divided by $\Delta\omega=2\pi/T_{tot}$, where $T_{tot}$ is a total time of our measurements (i.e. the difference between the time of the last and first samples in the data combined from all the laboratories). Equations (\ref{eq:CLX}) and (\ref{eq:DT}) are strictly valid for white noise. We tested with the Monte Carlo simulations, however, that Eqs. (\ref{eq:CLX}) and (\ref{eq:DT})  well approximate the confidence level and detection threshold for other types of noise; for this, one should replace the white noise parameter 
$\langle A^2 \rangle$ with the noise model function $\langle A^2 \rangle(\omega)$. For instance, in the case of pink noise, $\langle A^2 \rangle(\omega)\propto 1/\omega$, the $A^2$ corresponding to 5\% and 99\% CLs calculated as $A^2_X(\omega)=-\langle A^2\rangle (\omega) ln(1-X)$ is indistinguishable from accurate Monte Carlo simulations at the scale of Fig.~\ref{fig:osc}. A slight difference occurs for the detection threshold but it can be easily eliminated with a numerically determined correction factor   

\begin{equation}
A^2_{DT}(\omega)\approx-1.16\langle A^2\rangle (\omega)ln\left(\frac{0.05}{N_f}\right).
\end{equation}

We use the above expressions to determine the confidence levels and detection threshold.
%, see left panel in Fig.~\ref{fig:osc}. 
%In the frequency range from $\omega = 0.000043$~Hz to 0.032~Hz the noise can be modelled as a pink noise and for the frequencies higher than 0.032~Hz it can be modelled as a white noise.  
The periodogram in the left panel in Fig.~\ref{fig:osc} shows that in our measurements the power distribution in collected data is consistent with pink noise~\cite{Groth1975,Scargle1982-xm}.

\subsubsection{Alternative definition of the detection threshold}

Another way of defining the detection threshold is that $A^2$ evaluated from white noise does not exceed $A^2_{DT}$ in any of $N_f$ frequency channels at the confidence level $X$. It can be expressed as $X=[1-e^{-A^2_{DT}/\langle A^2\rangle}]^{N_f}$. Therefore, the detection threshold for a white noise, and for $N_f$ frequency channels with confidence level $X$ is

\begin{equation}
A^2_{DT}=-\langle A^2\rangle ln(1-X^{1/N_f}),
\end{equation}

\noindent
which for $X$ close to 1 can be written as 

\begin{equation}
A^2_{DT}\approx-\langle A^2\rangle ln\left(\frac{1-X}{N_f}\right).
\end{equation}

\noindent
This is equivalent to Eq. (\ref{eq:DT}).

\section*{Acknowledgements}

The "A next-generation worldwide quantum sensor network with optical atomic clocks" project is carried out within the TEAM IV programme of the Foundation for Polish Science co-financed by the European Union under the European Regional Development Fund. 
Support has been received from the project EMPIR 15SIB03 OC18. This project has received funding from the EMPIR programme co-financed by the Participating States and from the European Union's Horizon 2020 research and innovation programme. 
This project has received funding from NIST, NASA fundamental physics, and DARPA QuASAR.
We acknowledge funding support from the Agence Nationale de
la Recherche (Labex First-TF ANR-10-LABX-48-01), Centre National d'\'Etudes Spatiales (CNES), Conseil R\'egional \^{I}le-de-France (DIM Nano'K).
PW contribution is supported by the National Science Centre, Poland, Project no. 2015/19/D/ST2/02195 and the Polish Ministry of Science and Higher Education "Mobility Plus" Program. PM is the JSPS International Research Fellow.
MB contribution is supported by the National Science Centre, Poland under QuantERA programme no. 2017/25/Z/ST2/03021.
Calculations have been carried out using resources provided by the computer cluster founded by the Polish National Science Centre under Grant No. 2016/21/D/ST4/00903.

\bibliographystyle{apsrev4-1}

\begin{thebibliography}{40}%
\makeatletter
\providecommand \@ifxundefined [1]{%
 \@ifx{#1\undefined}
}%
\providecommand \@ifnum [1]{%
 \ifnum #1\expandafter \@firstoftwo
 \else \expandafter \@secondoftwo
 \fi
}%
\providecommand \@ifx [1]{%
 \ifx #1\expandafter \@firstoftwo
 \else \expandafter \@secondoftwo
 \fi
}%
\providecommand \natexlab [1]{#1}%
\providecommand \enquote  [1]{``#1''}%
\providecommand \bibnamefont  [1]{#1}%
\providecommand \bibfnamefont [1]{#1}%
\providecommand \citenamefont [1]{#1}%
\providecommand \href@noop [0]{\@secondoftwo}%
\providecommand \href [0]{\begingroup \@sanitize@url \@href}%
\providecommand \@href[1]{\@@startlink{#1}\@@href}%
\providecommand \@@href[1]{\endgroup#1\@@endlink}%
\providecommand \@sanitize@url [0]{\catcode `\\12\catcode `\$12\catcode
  `\&12\catcode `\#12\catcode `\^12\catcode `\_12\catcode `\%12\relax}%
\providecommand \@@startlink[1]{}%
\providecommand \@@endlink[0]{}%
\providecommand \url  [0]{\begingroup\@sanitize@url \@url }%
\providecommand \@url [1]{\endgroup\@href {#1}{\urlprefix }}%
\providecommand \urlprefix  [0]{URL }%
\providecommand \Eprint [0]{\href }%
\providecommand \doibase [0]{http://dx.doi.org/}%
\providecommand \selectlanguage [0]{\@gobble}%
\providecommand \bibinfo  [0]{\@secondoftwo}%
\providecommand \bibfield  [0]{\@secondoftwo}%
\providecommand \translation [1]{[#1]}%
\providecommand \BibitemOpen [0]{}%
\providecommand \bibitemStop [0]{}%
\providecommand \bibitemNoStop [0]{.\EOS\space}%
\providecommand \EOS [0]{\spacefactor3000\relax}%
\providecommand \BibitemShut  [1]{\csname bibitem#1\endcsname}%
\let\auto@bib@innerbib\@empty
%</preamble>
\bibitem [{\citenamefont {Wilczek}(1978)}]{Wilczek1978-va}%
  \BibitemOpen
  \bibfield  {author} {\bibinfo {author} {\bibfnamefont {F.}~\bibnamefont
  {Wilczek}},\ }\href@noop {} {\bibfield  {journal} {\bibinfo  {journal} {Phys.
  Rev. Lett.}\ }\textbf {\bibinfo {volume} {40}},\ \bibinfo {pages} {279}
  (\bibinfo {year} {1978})}\BibitemShut {NoStop}%
\bibitem [{\citenamefont {Weinberg}(1978)}]{Weinberg1978-qr}%
  \BibitemOpen
  \bibfield  {author} {\bibinfo {author} {\bibfnamefont {S.}~\bibnamefont
  {Weinberg}},\ }\href@noop {} {\bibfield  {journal} {\bibinfo  {journal}
  {Phys. Rev. Lett.}\ }\textbf {\bibinfo {volume} {40}},\ \bibinfo {pages}
  {223} (\bibinfo {year} {1978})}\BibitemShut {NoStop}%
\bibitem [{\citenamefont {Olive}(2016)}]{Olive2016-nh}%
  \BibitemOpen
  \bibfield  {author} {\bibinfo {author} {\bibfnamefont {K.~A.}\ \bibnamefont
  {Olive}},\ }\href@noop {} {\bibfield  {journal} {\bibinfo  {journal} {Chin.
  Phys. C}\ }\textbf {\bibinfo {volume} {40}},\ \bibinfo {pages} {100001}
  (\bibinfo {year} {2016})}\BibitemShut {NoStop}%
\bibitem [{\citenamefont {Feng}\ \emph {et~al.}(2003)\citenamefont {Feng},
  \citenamefont {Rajaraman},\ and\ \citenamefont {Takayama}}]{Feng2003-eh}%
  \BibitemOpen
  \bibfield  {author} {\bibinfo {author} {\bibfnamefont {J.~L.}\ \bibnamefont
  {Feng}}, \bibinfo {author} {\bibfnamefont {A.}~\bibnamefont {Rajaraman}}, \
  and\ \bibinfo {author} {\bibfnamefont {F.}~\bibnamefont {Takayama}},\
  }\href@noop {} {\bibfield  {journal} {\bibinfo  {journal} {Phys. Rev. Lett.}\
  }\textbf {\bibinfo {volume} {91}} (\bibinfo {year} {2003})}\BibitemShut
  {NoStop}%
\bibitem [{\citenamefont {Akerib}\ \emph {et~al.}(2014)\citenamefont {Akerib}
  \emph {et~al.}}]{noauthor_2014-um}%
  \BibitemOpen
  \bibfield  {author} {\bibinfo {author} {\bibfnamefont {D.~S.}\ \bibnamefont
  {Akerib}} \emph {et~al.} (\bibinfo {collaboration} {LUX Collaboration}),\
  }\href@noop {} {\bibfield  {journal} {\bibinfo  {journal} {Phys. Rev. Lett.}\
  }\textbf {\bibinfo {volume} {112}},\ \bibinfo {pages} {091303} (\bibinfo
  {year} {2014})}\BibitemShut {NoStop}%
\bibitem [{\citenamefont {Aprile}\ \emph {et~al.}(2017)\citenamefont {Aprile}
  \emph {et~al.}}]{noauthor_2017-zy}%
  \BibitemOpen
  \bibfield  {author} {\bibinfo {author} {\bibfnamefont {E.}~\bibnamefont
  {Aprile}} \emph {et~al.} (\bibinfo {collaboration} {XENON Collaboration}),\
  }\href@noop {} {\bibfield  {journal} {\bibinfo  {journal} {Phys. Rev. D}\
  }\textbf {\bibinfo {volume} {96}},\ \bibinfo {pages} {122002} (\bibinfo
  {year} {2017})}\BibitemShut {NoStop}%
\bibitem [{\citenamefont {Arvanitaki}\ \emph {et~al.}(2015)\citenamefont
  {Arvanitaki}, \citenamefont {Huang},\ and\ \citenamefont
  {Van~Tilburg}}]{Arvanitaki2015-bk}%
  \BibitemOpen
  \bibfield  {author} {\bibinfo {author} {\bibfnamefont {A.}~\bibnamefont
  {Arvanitaki}}, \bibinfo {author} {\bibfnamefont {J.}~\bibnamefont {Huang}}, \
  and\ \bibinfo {author} {\bibfnamefont {K.}~\bibnamefont {Van~Tilburg}},\
  }\href@noop {} {\bibfield  {journal} {\bibinfo  {journal} {Phys. Rev. D Part.
  Fields}\ }\textbf {\bibinfo {volume} {91}} (\bibinfo {year}
  {2015})}\BibitemShut {NoStop}%
\bibitem [{\citenamefont {Derevianko}\ and\ \citenamefont
  {Pospelov}(2014)}]{Derevianko2014-tz}%
  \BibitemOpen
  \bibfield  {author} {\bibinfo {author} {\bibfnamefont {A.}~\bibnamefont
  {Derevianko}}\ and\ \bibinfo {author} {\bibfnamefont {M.}~\bibnamefont
  {Pospelov}},\ }\href@noop {} {\bibfield  {journal} {\bibinfo  {journal} {Nat.
  Phys.}\ }\textbf {\bibinfo {volume} {10}},\ \bibinfo {pages} {933} (\bibinfo
  {year} {2014})}\BibitemShut {NoStop}%
\bibitem [{\citenamefont {Wcis{\l}o}\ \emph {et~al.}(2016)\citenamefont
  {Wcis{\l}o} \emph {et~al.}}]{Wcislo2016-cu}%
  \BibitemOpen
  \bibfield  {author} {\bibinfo {author} {\bibfnamefont {P.}~\bibnamefont
  {Wcis{\l}o}} \emph {et~al.},\ }\href@noop {} {\bibfield  {journal} {\bibinfo
  {journal} {Nature Astronomy}\ }\textbf {\bibinfo {volume} {1}},\ \bibinfo
  {pages} {0009} (\bibinfo {year} {2016})}\BibitemShut {NoStop}%
\bibitem [{\citenamefont {Damour}\ and\ \citenamefont
  {Donoghue}(2010{\natexlab{a}})}]{Damour2010-iu}%
  \BibitemOpen
  \bibfield  {author} {\bibinfo {author} {\bibfnamefont {T.}~\bibnamefont
  {Damour}}\ and\ \bibinfo {author} {\bibfnamefont {J.~F.}\ \bibnamefont
  {Donoghue}},\ }\href@noop {} {\bibfield  {journal} {\bibinfo  {journal}
  {Phys. Rev. D Part. Fields}\ }\textbf {\bibinfo {volume} {82}} (\bibinfo
  {year} {2010}{\natexlab{a}})}\BibitemShut {NoStop}%
\bibitem [{\citenamefont {Stadnik}\ and\ \citenamefont
  {Flambaum}(2016{\natexlab{a}})}]{Stadnik2016-ej}%
  \BibitemOpen
  \bibfield  {author} {\bibinfo {author} {\bibfnamefont {Y.~V.}\ \bibnamefont
  {Stadnik}}\ and\ \bibinfo {author} {\bibfnamefont {V.~V.}\ \bibnamefont
  {Flambaum}},\ }\href@noop {} {\bibfield  {journal} {\bibinfo  {journal}
  {Phys. Rev. A}\ }\textbf {\bibinfo {volume} {94}} (\bibinfo {year}
  {2016}{\natexlab{a}})}\BibitemShut {NoStop}%
\bibitem [{\citenamefont {Stadnik}\ and\ \citenamefont
  {Flambaum}(2014)}]{Stadnik2014-rx}%
  \BibitemOpen
  \bibfield  {author} {\bibinfo {author} {\bibfnamefont {Y.~V.}\ \bibnamefont
  {Stadnik}}\ and\ \bibinfo {author} {\bibfnamefont {V.~V.}\ \bibnamefont
  {Flambaum}},\ }\href@noop {} {\bibfield  {journal} {\bibinfo  {journal}
  {Phys. Rev. Lett.}\ }\textbf {\bibinfo {volume} {113}},\ \bibinfo {pages}
  {151301} (\bibinfo {year} {2014})}\BibitemShut {NoStop}%
\bibitem [{\citenamefont {Hees}\ \emph {et~al.}(2016)\citenamefont {Hees},
  \citenamefont {Gu{\'e}na}, \citenamefont {Abgrall}, \citenamefont {Bize},\
  and\ \citenamefont {Wolf}}]{Hees2016-cf}%
  \BibitemOpen
  \bibfield  {author} {\bibinfo {author} {\bibfnamefont {A.}~\bibnamefont
  {Hees}}, \bibinfo {author} {\bibfnamefont {J.}~\bibnamefont {Gu{\'e}na}},
  \bibinfo {author} {\bibfnamefont {M.}~\bibnamefont {Abgrall}}, \bibinfo
  {author} {\bibfnamefont {S.}~\bibnamefont {Bize}}, \ and\ \bibinfo {author}
  {\bibfnamefont {P.}~\bibnamefont {Wolf}},\ }\href@noop {} {\bibfield
  {journal} {\bibinfo  {journal} {Phys. Rev. Lett.}\ }\textbf {\bibinfo
  {volume} {117}} (\bibinfo {year} {2016})}\BibitemShut {NoStop}%
\bibitem [{\citenamefont {Van~Tilburg}\ \emph {et~al.}(2015)\citenamefont
  {Van~Tilburg}, \citenamefont {Leefer}, \citenamefont {Bougas},\ and\
  \citenamefont {Budker}}]{Van_Tilburg2015-uf}%
  \BibitemOpen
  \bibfield  {author} {\bibinfo {author} {\bibfnamefont {K.}~\bibnamefont
  {Van~Tilburg}}, \bibinfo {author} {\bibfnamefont {N.}~\bibnamefont {Leefer}},
  \bibinfo {author} {\bibfnamefont {L.}~\bibnamefont {Bougas}}, \ and\ \bibinfo
  {author} {\bibfnamefont {D.}~\bibnamefont {Budker}},\ }\href@noop {}
  {\bibfield  {journal} {\bibinfo  {journal} {Phys. Rev. Lett.}\ }\textbf
  {\bibinfo {volume} {115}},\ \bibinfo {pages} {011802} (\bibinfo {year}
  {2015})}\BibitemShut {NoStop}%
\bibitem [{\citenamefont {Kalaydzhyan}\ and\ \citenamefont
  {Yu}(2017)}]{Kalaydzhyan2017-wi}%
  \BibitemOpen
  \bibfield  {author} {\bibinfo {author} {\bibfnamefont {T.}~\bibnamefont
  {Kalaydzhyan}}\ and\ \bibinfo {author} {\bibfnamefont {N.}~\bibnamefont
  {Yu}},\ }\href@noop {} {\bibfield  {journal} {\bibinfo  {journal} {Phys. Rev.
  D Part. Fields}\ }\textbf {\bibinfo {volume} {96}} (\bibinfo {year}
  {2017})}\BibitemShut {NoStop}%
\bibitem [{\citenamefont {Roberts}\ \emph {et~al.}(2017)\citenamefont {Roberts}
  \emph {et~al.}}]{Roberts2017-xk}%
  \BibitemOpen
  \bibfield  {author} {\bibinfo {author} {\bibfnamefont {B.~M.}\ \bibnamefont
  {Roberts}} \emph {et~al.},\ }\href@noop {} {\bibfield  {journal} {\bibinfo
  {journal} {Nat. Commun.}\ }\textbf {\bibinfo {volume} {8}},\ \bibinfo {pages}
  {1195} (\bibinfo {year} {2017})}\BibitemShut {NoStop}%
\bibitem [{\citenamefont {Blatt}\ \emph {et~al.}(2008)\citenamefont {Blatt}
  \emph {et~al.}}]{Blatt2008}%
  \BibitemOpen
  \bibfield  {author} {\bibinfo {author} {\bibfnamefont {S.}~\bibnamefont
  {Blatt}} \emph {et~al.},\ }\href {\doibase 10.1103/PhysRevLett.100.140801}
  {\bibfield  {journal} {\bibinfo  {journal} {Phys. Rev. Lett.}\ }\textbf
  {\bibinfo {volume} {100}},\ \bibinfo {pages} {140801} (\bibinfo {year}
  {2008})}\BibitemShut {NoStop}%
\bibitem [{\citenamefont {Delva}\ \emph {et~al.}(2017)\citenamefont {Delva}
  \emph {et~al.}}]{Delva2017}%
  \BibitemOpen
  \bibfield  {author} {\bibinfo {author} {\bibfnamefont {P.}~\bibnamefont
  {Delva}} \emph {et~al.},\ }\href {\doibase 10.1103/PhysRevLett.118.221102}
  {\bibfield  {journal} {\bibinfo  {journal} {Phys. Rev. Lett.}\ }\textbf
  {\bibinfo {volume} {118}},\ \bibinfo {pages} {221102} (\bibinfo {year}
  {2017})}\BibitemShut {NoStop}%
\bibitem [{\citenamefont {Budker}\ and\ \citenamefont
  {Derevianko}(2015)}]{Budker2015}%
  \BibitemOpen
  \bibfield  {author} {\bibinfo {author} {\bibfnamefont {D.}~\bibnamefont
  {Budker}}\ and\ \bibinfo {author} {\bibfnamefont {A.}~\bibnamefont
  {Derevianko}},\ }\href {\doibase 10.1063/PT.3.2896} {\bibfield  {journal}
  {\bibinfo  {journal} {Physics Today}\ }\textbf {\bibinfo {volume} {68}},\
  \bibinfo {pages} {10} (\bibinfo {year} {2015})}\BibitemShut {NoStop}%
\bibitem [{\citenamefont {Schioppo}\ \emph {et~al.}(2016)\citenamefont
  {Schioppo} \emph {et~al.}}]{Schioppo2016-va}%
  \BibitemOpen
  \bibfield  {author} {\bibinfo {author} {\bibfnamefont {M.}~\bibnamefont
  {Schioppo}} \emph {et~al.},\ }\href@noop {} {\bibfield  {journal} {\bibinfo
  {journal} {Nat. Photonics}\ }\textbf {\bibinfo {volume} {11}},\ \bibinfo
  {pages} {48} (\bibinfo {year} {2016})}\BibitemShut {NoStop}%
\bibitem [{\citenamefont {Hinkley}\ \emph {et~al.}(2013)\citenamefont {Hinkley}
  \emph {et~al.}}]{Hinkley2013-su}%
  \BibitemOpen
  \bibfield  {author} {\bibinfo {author} {\bibfnamefont {N.}~\bibnamefont
  {Hinkley}} \emph {et~al.},\ }\href@noop {} {\bibfield  {journal} {\bibinfo
  {journal} {Science}\ }\textbf {\bibinfo {volume} {341}},\ \bibinfo {pages}
  {1215} (\bibinfo {year} {2013})}\BibitemShut {NoStop}%
\bibitem [{\citenamefont {Le~Targat}\ \emph {et~al.}(2013)\citenamefont
  {Le~Targat} \emph {et~al.}}]{Le_Targat2013-eu}%
  \BibitemOpen
  \bibfield  {author} {\bibinfo {author} {\bibfnamefont {R.}~\bibnamefont
  {Le~Targat}} \emph {et~al.},\ }\href@noop {} {\bibfield  {journal} {\bibinfo
  {journal} {Nat. Commun.}\ }\textbf {\bibinfo {volume} {4}},\ \bibinfo {pages}
  {2109} (\bibinfo {year} {2013})}\BibitemShut {NoStop}%
\bibitem [{\citenamefont {Lodewyck}\ \emph {et~al.}(2016)\citenamefont
  {Lodewyck} \emph {et~al.}}]{Lodewyck2016-pk}%
  \BibitemOpen
  \bibfield  {author} {\bibinfo {author} {\bibfnamefont {J.}~\bibnamefont
  {Lodewyck}} \emph {et~al.},\ }\href@noop {} {\bibfield  {journal} {\bibinfo
  {journal} {Metrologia}\ }\textbf {\bibinfo {volume} {53}},\ \bibinfo {pages}
  {1123} (\bibinfo {year} {2016})}\BibitemShut {NoStop}%
\bibitem [{\citenamefont {Morzy{\'n}ski}\ \emph {et~al.}(2015)\citenamefont
  {Morzy{\'n}ski} \emph {et~al.}}]{Morzynski2015-yn}%
  \BibitemOpen
  \bibfield  {author} {\bibinfo {author} {\bibfnamefont {P.}~\bibnamefont
  {Morzy{\'n}ski}} \emph {et~al.},\ }\href@noop {} {\bibfield  {journal}
  {\bibinfo  {journal} {Sci. Rep.}\ }\textbf {\bibinfo {volume} {5}},\ \bibinfo
  {pages} {17495} (\bibinfo {year} {2015})}\BibitemShut {NoStop}%
\bibitem [{\citenamefont {Bober}\ \emph {et~al.}(2015)\citenamefont {Bober}
  \emph {et~al.}}]{Bober2015-up}%
  \BibitemOpen
  \bibfield  {author} {\bibinfo {author} {\bibfnamefont {M.}~\bibnamefont
  {Bober}} \emph {et~al.},\ }\href@noop {} {\bibfield  {journal} {\bibinfo
  {journal} {Measurement Science and Technology}\ }\textbf {\bibinfo {volume}
  {26}},\ \bibinfo {pages} {075201} (\bibinfo {year} {2015})}\BibitemShut
  {NoStop}%
\bibitem [{\citenamefont {Hachisu}\ and\ \citenamefont
  {Ido}(2015)}]{Hachisu2015-kq}%
  \BibitemOpen
  \bibfield  {author} {\bibinfo {author} {\bibfnamefont {H.}~\bibnamefont
  {Hachisu}}\ and\ \bibinfo {author} {\bibfnamefont {T.}~\bibnamefont {Ido}},\
  }\href@noop {} {\bibfield  {journal} {\bibinfo  {journal} {Jpn. J. Appl.
  Phys.}\ }\textbf {\bibinfo {volume} {54}},\ \bibinfo {pages} {112401}
  (\bibinfo {year} {2015})}\BibitemShut {NoStop}%
\bibitem [{\citenamefont {Hachisu}\ \emph {et~al.}(2018)\citenamefont
  {Hachisu}, \citenamefont {Nakagawa}, \citenamefont {Hanado},\ and\
  \citenamefont {Ido}}]{Hachisu2018}%
  \BibitemOpen
  \bibfield  {author} {\bibinfo {author} {\bibfnamefont {H.}~\bibnamefont
  {Hachisu}}, \bibinfo {author} {\bibfnamefont {F.}~\bibnamefont {Nakagawa}},
  \bibinfo {author} {\bibfnamefont {Y.}~\bibnamefont {Hanado}}, \ and\ \bibinfo
  {author} {\bibfnamefont {T.}~\bibnamefont {Ido}},\ }\href@noop {} {\bibfield
  {journal} {\bibinfo  {journal} {Scientific Reports}\ }\textbf {\bibinfo
  {volume} {8}},\ \bibinfo {pages} {4243} (\bibinfo {year} {2018})}\BibitemShut
  {NoStop}%
\bibitem [{\citenamefont {Stadnik}\ and\ \citenamefont
  {Flambaum}(2015{\natexlab{a}})}]{Stadnik2015-cq}%
  \BibitemOpen
  \bibfield  {author} {\bibinfo {author} {\bibfnamefont {Y.~V.}\ \bibnamefont
  {Stadnik}}\ and\ \bibinfo {author} {\bibfnamefont {V.~V.}\ \bibnamefont
  {Flambaum}},\ }\href@noop {} {\bibfield  {journal} {\bibinfo  {journal}
  {Phys. Rev. Lett.}\ }\textbf {\bibinfo {volume} {114}} (\bibinfo {year}
  {2015}{\natexlab{a}})}\BibitemShut {NoStop}%
\bibitem [{\citenamefont {Stadnik}\ and\ \citenamefont
  {Flambaum}(2016{\natexlab{b}})}]{Stadnik2016-cm}%
  \BibitemOpen
  \bibfield  {author} {\bibinfo {author} {\bibfnamefont {Y.~V.}\ \bibnamefont
  {Stadnik}}\ and\ \bibinfo {author} {\bibfnamefont {V.~V.}\ \bibnamefont
  {Flambaum}},\ }\href@noop {} {\bibfield  {journal} {\bibinfo  {journal}
  {Phys. Rev. A}\ }\textbf {\bibinfo {volume} {93}} (\bibinfo {year}
  {2016}{\natexlab{b}})}\BibitemShut {NoStop}%
\bibitem [{\citenamefont {Essig}\ \emph {et~al.}(2013)\citenamefont {Essig}
  \emph
  {et~al.}}]{Essig_Rouven_and_Jaros_John_A_and_Wester_William_et_al2013-mm}%
  \BibitemOpen
  \bibfield  {author} {\bibinfo {author} {\bibfnamefont {R.}~\bibnamefont
  {Essig}} \emph {et~al.},\ }\href@noop {} {\bibfield  {journal} {\bibinfo
  {journal} {arXiv}\ ,\ \bibinfo {pages} {1311.0029}} (\bibinfo {year}
  {2013})}\BibitemShut {NoStop}%
\bibitem [{\citenamefont {Vilenkin}(1985)}]{Vilenkin1985-lb}%
  \BibitemOpen
  \bibfield  {author} {\bibinfo {author} {\bibfnamefont {A.}~\bibnamefont
  {Vilenkin}},\ }\href@noop {} {\bibfield  {journal} {\bibinfo  {journal}
  {Phys. Rep.}\ }\textbf {\bibinfo {volume} {121}},\ \bibinfo {pages} {263}
  (\bibinfo {year} {1985})}\BibitemShut {NoStop}%
\bibitem [{\citenamefont {Derevianko}(2016)}]{Derevianko2016-pq}%
  \BibitemOpen
  \bibfield  {author} {\bibinfo {author} {\bibfnamefont {A.}~\bibnamefont
  {Derevianko}},\ }\href@noop {} {\bibfield  {journal} {\bibinfo  {journal} {J.
  Phys. Conf. Ser.}\ }\textbf {\bibinfo {volume} {723}},\ \bibinfo {pages}
  {012043} (\bibinfo {year} {2016})}\BibitemShut {NoStop}%
\bibitem [{\citenamefont {Damour}\ and\ \citenamefont
  {Donoghue}(2010{\natexlab{b}})}]{Damour2010-mr}%
  \BibitemOpen
  \bibfield  {author} {\bibinfo {author} {\bibfnamefont {T.}~\bibnamefont
  {Damour}}\ and\ \bibinfo {author} {\bibfnamefont {J.~F.}\ \bibnamefont
  {Donoghue}},\ }\href@noop {} {\bibfield  {journal} {\bibinfo  {journal}
  {Classical Quantum Gravity}\ }\textbf {\bibinfo {volume} {27}},\ \bibinfo
  {pages} {202001} (\bibinfo {year} {2010}{\natexlab{b}})}\BibitemShut
  {NoStop}%
\bibitem [{\citenamefont {Stadnik}\ and\ \citenamefont
  {Flambaum}(2015{\natexlab{b}})}]{Stadnik2015-ho}%
  \BibitemOpen
  \bibfield  {author} {\bibinfo {author} {\bibfnamefont {Y.~V.}\ \bibnamefont
  {Stadnik}}\ and\ \bibinfo {author} {\bibfnamefont {V.~V.}\ \bibnamefont
  {Flambaum}},\ }\href@noop {} {\bibfield  {journal} {\bibinfo  {journal}
  {Phys. Rev. Lett.}\ }\textbf {\bibinfo {volume} {115}},\ \bibinfo {pages}
  {201301} (\bibinfo {year} {2015}{\natexlab{b}})}\BibitemShut {NoStop}%
\bibitem [{\citenamefont {{Ya. B. Zel'dovich, I. Yu. Kobzarev, L.B.
  Okun'}}(1975)}]{Ya_B_Zeldovich_I_Yu_Kobzarev_LB_Okun1975-ju}%
  \BibitemOpen
  \bibfield  {author} {\bibinfo {author} {\bibnamefont {{Ya. B. Zel'dovich, I.
  Yu. Kobzarev, L.B. Okun'}}},\ }\href@noop {} {\bibfield  {journal} {\bibinfo
  {journal} {JETP}\ }\textbf {\bibinfo {volume} {40}},\ \bibinfo {pages} {1}
  (\bibinfo {year} {1975})}\BibitemShut {NoStop}%
\bibitem [{\citenamefont {Scargle}(1982)}]{Scargle1982-xm}%
  \BibitemOpen
  \bibfield  {author} {\bibinfo {author} {\bibfnamefont {J.~D.}\ \bibnamefont
  {Scargle}},\ }\href@noop {} {\bibfield  {journal} {\bibinfo  {journal} {ApJ}\
  }\textbf {\bibinfo {volume} {263}},\ \bibinfo {pages} {835} (\bibinfo {year}
  {1982})}\BibitemShut {NoStop}%
\bibitem [{\citenamefont {{Groth}}(1975)}]{Groth1975}%
  \BibitemOpen
  \bibfield  {author} {\bibinfo {author} {\bibfnamefont {E.~J.}\ \bibnamefont
  {{Groth}}},\ }\href {\doibase 10.1086/190343} {\bibfield  {journal} {\bibinfo
   {journal} {ApJS}\ }\textbf {\bibinfo {volume} {29}},\ \bibinfo {pages} {285}
  (\bibinfo {year} {1975})}\BibitemShut {NoStop}%
\bibitem [{\citenamefont {Schlamminger}\ \emph {et~al.}(2008)\citenamefont
  {Schlamminger}, \citenamefont {Choi}, \citenamefont {Wagner}, \citenamefont
  {Gundlach},\ and\ \citenamefont {Adelberger}}]{Schlamminger2008-pf}%
  \BibitemOpen
  \bibfield  {author} {\bibinfo {author} {\bibfnamefont {S.}~\bibnamefont
  {Schlamminger}}, \bibinfo {author} {\bibfnamefont {K.-Y.}\ \bibnamefont
  {Choi}}, \bibinfo {author} {\bibfnamefont {T.~A.}\ \bibnamefont {Wagner}},
  \bibinfo {author} {\bibfnamefont {J.~H.}\ \bibnamefont {Gundlach}}, \ and\
  \bibinfo {author} {\bibfnamefont {E.~G.}\ \bibnamefont {Adelberger}},\
  }\href@noop {} {\bibfield  {journal} {\bibinfo  {journal} {Phys. Rev. Lett.}\
  }\textbf {\bibinfo {volume} {100}} (\bibinfo {year} {2008})}\BibitemShut
  {NoStop}%
\bibitem [{\citenamefont {Williams}\ \emph {et~al.}(2004)\citenamefont
  {Williams}, \citenamefont {Turyshev},\ and\ \citenamefont
  {Boggs}}]{Williams2004-ou}%
  \BibitemOpen
  \bibfield  {author} {\bibinfo {author} {\bibfnamefont {J.~G.}\ \bibnamefont
  {Williams}}, \bibinfo {author} {\bibfnamefont {S.~G.}\ \bibnamefont
  {Turyshev}}, \ and\ \bibinfo {author} {\bibfnamefont {D.~H.}\ \bibnamefont
  {Boggs}},\ }\href@noop {} {\bibfield  {journal} {\bibinfo  {journal} {Phys.
  Rev. Lett.}\ }\textbf {\bibinfo {volume} {93}} (\bibinfo {year}
  {2004})}\BibitemShut {NoStop}%
\bibitem [{\citenamefont {Berg\'e}\ \emph {et~al.}(2018)\citenamefont
  {Berg\'e}, \citenamefont {Brax}, \citenamefont {M\'etris}, \citenamefont
  {Pernot-Borr\`as}, \citenamefont {Touboul},\ and\ \citenamefont
  {Uzan}}]{Berge2017-ys}%
  \BibitemOpen
  \bibfield  {author} {\bibinfo {author} {\bibfnamefont {J.}~\bibnamefont
  {Berg\'e}}, \bibinfo {author} {\bibfnamefont {P.}~\bibnamefont {Brax}},
  \bibinfo {author} {\bibfnamefont {G.}~\bibnamefont {M\'etris}}, \bibinfo
  {author} {\bibfnamefont {M.}~\bibnamefont {Pernot-Borr\`as}}, \bibinfo
  {author} {\bibfnamefont {P.}~\bibnamefont {Touboul}}, \ and\ \bibinfo
  {author} {\bibfnamefont {J.-P.}\ \bibnamefont {Uzan}},\ }\href {\doibase
  10.1103/PhysRevLett.120.141101} {\bibfield  {journal} {\bibinfo  {journal}
  {Phys. Rev. Lett.}\ }\textbf {\bibinfo {volume} {120}},\ \bibinfo {pages}
  {141101} (\bibinfo {year} {2018})}\BibitemShut {NoStop}%
\end{thebibliography}

\end{document}